\documentstyle[aps,prd,epsfig,floats]{revtex} 

\draft

\begin{document}
\twocolumn[\hsize\textwidth\columnwidth\hsize\csname
@twocolumnfalse\endcsname
\title{%
\hbox to\hsize{\normalsize\rm April 2000
\hfil Preprint MPI-Pth/2000-15}
\vskip 36pt Study of Macroscopic Coherence  and Decoherence in the
SQUID by Adiabatic
Inversion}

\author{P. Silvestrini}
\address{Instituto di Cibernetica del CNR, via Toiano 6, I-
80072, Arco Felice, Italy and
MQC group, INFN/Napoli, Italy}

\author{L.~Stodolsky}
\address{Max-Planck-Institut f\"ur Physik 
(Werner-Heisenberg-Institut),
F\"ohringer Ring 6, 80805 M\"unchen, Germany}

\date{April 2000}
\maketitle
\begin{abstract}
 We suggest a procedure
for demonstrating quantum coherence and measuring decoherence times
 between different
fluxoid states of a SQUID by using  ``adiabatic
inversion'', where one macroscopic fluxoid state is smoothly
transferred into
the other,  like a spin  reversing direction by following  a
slowly
moving
 magnetic field. This is accomplished by sweeping an external
applied flux, and
  depends on a well-defined quantum
phase between the two macroscopic states.  Varying the speed
of the sweep relative to the decoherence time permits one to
move from the quantum regime, where such a well-defined phase
exists, to the classical regime where it is lost and
the inversion is inhibited. Thus  observing 
whether inversion
has taken place or not as a function of sweep speed
 offers the possibility of measuring
the  decoherence time.   The main
 requirement for the feasibility of the scheme appears to be that
the low
temperature relaxation
time among the quantum levels of the SQUID be long compared to
other  time scales of the problem. Applications to
 the ``quantum computer'', with the  level system
 of the SQUID playing the role  of the qbit, are briefly discussed.
\end{abstract}
\vskip2.0pc]

\section{Introduction}

One of  the most interesting questions of ``macroscopic quantum
mechanics'' 
concerns the
demonstration of  quantum coherence and interference
effects between apparently grossly distinguishable states of
 large systems. Observation of such effects would definitively lay
to rest any ideas about big or ``classical'' systems being
fundamentally different than small systems, in some way not subject
to the rules of quantum mechanics. One system which has been
extensively discussed in this regard is the Josephson
junction~\cite{ajl}
under
current bias and the closely related rf Squid, where a
superconducting ring is
interrupted
by a  Josephson junction. In these systems the phase across the
junction or the
flux  $\Phi$ in the ring plays the role of
a
collective coordinate. In the last decade a number of beautiful
junction experiments at
low temperature ~\cite{lukens} have seen effects connected with the
quantized  energy levels~\cite{clarke}
expected with respect to  this coordinate.
A fast sweeping
method ~\cite{obs} has also
seen the effects of these quantized levels even at relatively high
temperature. 

An even more striking and direct manifestation of macroscopic
quantum behavior would be
the demonstration of quantum coherence between two apparently
macroscopically different states of the rf SQUID, 
between
 states where the current goes   around the
ring in opposite directions. Recently, results have been reported
with microwave methods showing the repulsion of energy levels
expected due to  quantum
mechanical mixing of the different fluxoid states \cite{luk1}. Here
we would like to suggest a new method for demonstrating 
macroscopic coherence,
 one which further allows a measurement of the decoherence time,
the
 time in which the coherence between the states is lost.
 Our basic idea  is to use the process of  
``adiabatic inversion'' where a slowly varying field is used to
reverse the states of a quantum
system.
In its most straightforward realization, our proposal consists of
starting with the system in its lowest state, making a fast but
adiabatic sweep, and reading out to see if the final state is the
same or the opposite fluxoid state. If the state has switched, the
system has behaved quantum mechanically, with phase coherence
between the two states. If not, the phase coherence between the
states was lost and the system stayed in its original
configuration--
 behaved classically.

A very interesting aspect of the present proposal, as we shall
discuss
below, is the possiblity of passing from the 
quantum to the classical regime by simply varying 
the sweep speed. This allows us to obtain, for adiabatic
conditions,  the decoherence
time as the longest sweep time for which
the inversion is successful.

\section{Adiabatic Inversion in the rf Squid}

For the rf SQUID biased with an external flux $\Phi_x$,
the  ``potential energy'' of the system  for which $\Phi$ is the
coordinate is  shown in  Fig 1. One has, near the origin,  the 
 familiar  double  well potential ~\cite{barone}
\begin{equation}
\label{u}
U=U_0[1/2 (\Phi-\Phi_x)^2 + \beta_L cos\Phi]\;.
\end{equation} Depending on whether the system is in the left or
right well,
the flux through the ring has a different sign and  the current
goes around
 in opposite directions.
   The potential can be varied by altering the
external $\Phi_x$,  becoming 
symmetric for $\Phi_x=0$. We suppose a situation where a few states
may be
 possible in a well, but we will be essentially  concerned with
the
lowest level in each. In the adiabatic inversion procedure to be
discussed,
 $\Phi_x$ is swept from a maximum to an minimum value, passing
through zero, such that the
initially
asymmetric left and right wells exchange roles, the
originally higher
well becoming the lower one and vice versa. The asymmetry of the
configurations,
 however, is kept small so that we effectively have only
a two-state 
system, composed of the lowest state in each well. 

\begin{figure}[h]
\epsfig{file=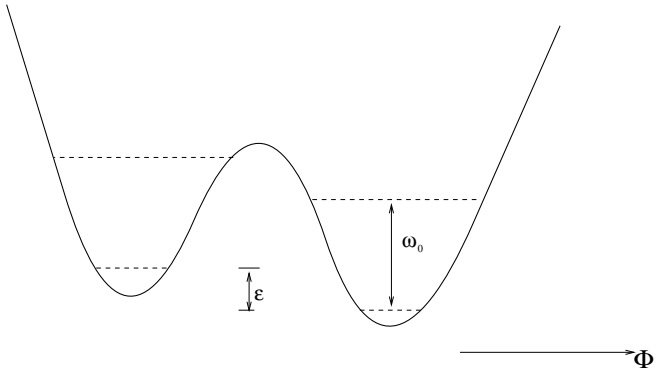, width=\hsize}
\caption{ Double potential well with harmonic level spacing 
$\omega_0$ and initial spacing $\epsilon$ between lowest levels.}
\end{figure}

 As quantum states we
refer to  the lowest state of the  left or right well as
$\vert L>$ and $\vert R>$, and the 
 tunneling between $\vert L>$ and $\vert R>$ will generally lead
to eigenstates which are linear combinations of the two.
Any
  two-state system  may viewed as constituting the two
components of
a ``spin'', and this provides an easy visualization which has been
used in many contexts~\cite{us}.   In the present problem ``spin
up'' can be identified with, say, the state $\vert L >$  (flux
one way) and ``spin
down'' with $\vert R >$ (flux the other way), while the spin in
a general direction
represents a quantum mechanical linear combination of the two
states, with some definite relative  magnitude and  phase. 
Various influences like the external
bias or
the  tunneling amplitude  may be thought of
as creating a kind of  pseudo-magnetic field $\bf V$ about which
the spin or polarization $\bf P$ 
precesses (Fig. 2) according to the equation

\begin{equation}\label{pdot}
{\bf \dot P} = {\bf V} \times {\bf P} -D {\bf P}_{tr}\; ,
\end{equation}
where $\bf V$ can be time dependent. The quantity $D$ is the
decoherence parameter, which 
 we  neglect for the moment and will  deal with below.

 Now it is a familiar fact  under adiabatic conditions, where
$\bf V$
varies slowly, that the  ``spin'' $\bf P$
will tend to ``follow'' the magnetic field ${\bf V}(t)$.  
 This is a completely familiar procedure  when rotating the spin
of say an atom or
 a neutron by a magnetic field.
 Here the  motion of $\bf V$ will be accomplished by sweeping
$\Phi_x$;  it is then merely necessary to identify
the components of $\bf V$  and the meaning of ``adiabatic'' or
``slow''.

\begin{figure}[h] 
\epsfig{file=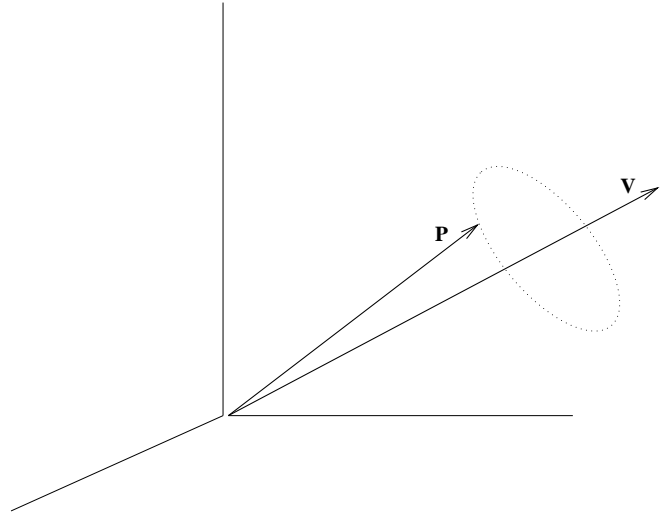, width=\hsize}
\caption{ Precession of the ``spin'' vector $\bf P$
 around the pseudo-magnetic field vector ${\bf V}(t)$. In adiabatic
inversion $\bf V$,  swings from ``up'' to
``down'', and carries $\bf P$ with it.}
\end{figure}
 The vertical component of $\bf V$
corresponds to the  difference in the two potential wells or lowest
energy
levels, for small asymmetry it is linear in $\Phi_x$. If $\epsilon$
is the initial level splitting when $\Phi_x=\Phi_x^{max}$, we can
write
$V_{vert}(t)=\epsilon (\Phi_x(t)/\Phi_x^{max})$. Thus as $\Phi_x$
sweeps from its positive maximum value to its negative minimum
value
$V_{vert}$ reverses direction.
 The
transverse component of ${\bf V}_{tr}$ corresponds to the tunneling
energy
between
the two quasi-degenerate states $V_{tr}=\omega_{tunnel}$. As
$V_{vert}$  passes through zero at 
$\Phi_x=0$, the $\vert L>$ and $\vert R>$ states are strongly mixed
and the splitting of the resulting energy eigenstates is determined
by  $V_{tr}$ alone.
 The magnitude of $V(t)$ at a given time,  
$|{\bf V}|=\sqrt{V_{vert}^2+V_{tr}^2}$  gives the instantaneous
splitting of the two levels, which varies from approximately
$\epsilon$ in the vertical position of $\bf V$ to
$\omega_{tunnel}$ in the horizontal position.  

As to the meaning of
``adiabatic'' or ``slow'',  the imposed time 
variation of $\bf V$ should  not significantly
contain frequencies or fourier components corresponding to the
energy splitting between levels.  Or in terms of time, the rate of
variation of  $\bf V$
should be slow on the time scale corresponding to the tunneling
time between the two states. Thus we have the requirement on $\bf
V$ that its relative rate of variation $\dot V/V$ always be small
compared to $ V$ itself. Since the varying component of $\bf V$ is
$V_{vertical}$ (neglecting the indirect effect of $\Phi_x$ on
$\omega_{tunnel}$) we require  $\dot V_{vertical}/V<<V $. Thus
taking the near degenerate configuration where $V\approx
\omega_{tunnel} $ we find  
\begin{equation}\label{adb}
 \epsilon {\dot \Phi_x(t)\over \Phi_x^{max}}\approx \epsilon
\omega_{sweep} << \omega_{tunnel}^2
\end{equation}
 as the
condition
 for adiabaticity. If the adiabatic condition is violated, then
$\bf P$ cannot follow $\bf V $ and undergoes wide swings.
We stress that for adiabatic inversion there must be a
 well-defined quantum phase between the two states, and that when
this
phase is lost the inversion is suppressed. This is the topic of the
next section.
\section{Damping} Thus far we have not
considered dissipative or damping effects
tending to destroy the quantum coherence of the system;
 we have neglected the $D$ term in Eq [\ref{pdot}]. The quantity
${\bf P}_{tr}$
has the interpretation of the degree of phase coherence between the
two states,
and $D$ gives the rate of loss of this coherence \cite{d}.
In particular,
 we are interested in the effect of $D$ on the  inversion process.
 With the loss of phase
coherence between the
two states, we expect the situation to become more ``classical''
and for the   inversion to be inhibited. Indeed, in solving
Eq[\ref{pdot}] for large $D$ one finds  that the inversion is
strongly blocked and  that
 one arrives in the ``Turing-Watched Pot-Zeno'' regime~\cite{us}.
 Eq [\ref{pdot}] with  moving $\bf V(t)$ has been studied
numerically~\cite{flaig}.
 One finds that
while for weak damping $D$ has little
effect on the
inversion,
 values of $D/V_{tr}\sim 0.1$ are enough to stall the inversion and
values 
$D/V_{tr}>>1$ block it altogether. Given adiabatic conditions, the
important
relation is that between the sweep speed
and the decoherence time $1/D$. This relation  determines if the
system has
time to ``decohere'' during the sweep. For $D/V_{tr}=0.1$, for
example, one finds
that a factor of ten increase in the sweep speed will change  the
stalled inversion into a successful one. Such behavior
 is of great interest to us here,
 since  if the experimental situation is
favorable,  we can  pass from the quantum regime (small $D$) to
the classical regime
(large $D$) by varying the sweep speed. In the former case the
system ends
up in the other potential well while in the
latter
case, where  the phase coherence is destroyed, the system behaves
 classically as ${\bf P}_{tr}\rightarrow 0$ and remains in the
original state.

 Concerning the absolute magnitude of $D$, a number of subtle
issues are involved which cannot be treated here. However, for
orientation we can take the result of calculations based on the
Caldeira-Leggett approach, where we may identify the ``decay rate''
for weak dissipation with  $D$. Weiss and Grabert~\cite{wg} give
 $T/Re^2$ for this parameter (their $\Gamma$). 
With
$R=5~M\Omega$, this gives $D= 0.08~mk=9.6~MHZ$  at $T=100~mk$, and
$D=
0.008~mk=960~kHZ$  for $T=10~mK$. (Units: $1~K= 8~meV=120~GHZ$,
$1/e^2=4~k\Omega$). If these estimates are correct,
we are
in an interesting range  since it is possible to suppose sweeps
 both slow and fast on these scales and thus the direct
measurement of $D$ and its temperature dependence.

\section{Time Scales}
Since we envision working at time scales shorter than have been
common
in this field, 
it may be useful to give a qualitative discussion of the various
time scales which arise. 
 The highest
frequency involved is the ordinary harmonic frequency
$\omega_0$ in
one of the potential wells,  giving
 the approximate level spacing 
  in the well. For typical conditions this spacing may be on
the order
of
 several Kelvin  ($K$). Since we
envision
experiments in the Kelvin to milliKelvin range, this implies that
for the equilibrium
system at the start of the sweep
 only the lowest level is populated.
 The next parameter is the tunneling
frequency through the barrier.
 This is a sensitive function of the SQUID
parameters, with sample values [$\beta_L= 1.8,~ C=.01~pF,
 L= 1~Nh, ~ R=5~M\Omega$] we obtain $ \approx 20~GHZ$, and 
$\omega_{tunnel}/\omega_{0}\approx 4\cdot 10^{-2}$.

 The
 next two parameters concern the extent and speed of the sweep. 
 The initial asymmetry  $\epsilon$ (Fig. 1) should
be small compared to $\omega_{0}$ in order to retain
 the approximate two-state character of the system, but large
compared to
 $\omega_{tunnel}$ to avoid initial mixing of  the two states,
 say $\epsilon/ \omega_{0}\sim 10^{-1}-10^{-2}$. The speed of the
sweep,
 $\omega_{sweep}$ will be the easiest  experimental
 parameter to control, and the behavior of the results as
$\omega_{sweep}$ is varied will be an 
important check on the theory. It must not be so fast
as to
lose adiabaticity, but not so slow as to allow
relaxation processes to mask
the results. We may suppose it to be in the range
 $10^{-3}\omega_{0}-10^{-4}\omega_{0}= 10-100~MHZ $. 
With these numbers, it is possible to satisfy the
adiabatic condition Eq~[\ref{adb}].

 Finally there are
dissipative parameters $D$,
 and $\omega_{relax}$, the
relaxation
rate for transitions among the SQUID levels.  From the above
estimate of $1/D$
in the 1-10~MHZ range we have, as mentioned, the
interesting
possiblity of being able to choose  sweep speeds either slow or
fast
with respect to the decoherence time while still retaining the
adiabatic condition.
The resulting switches between
classical and  quantum behavior would provide persuasive evidence
for the correctness of our general picture,
 and allow the measurement of this parameter and  its temperature
dependence.   Finally, the
relaxation
 parameter  $\omega_{relax}$ gives the rate of
conventional kinetic
 relaxation processes  induced by the 
environment,
 say  thermal jumping over the barrier
or  direct transitions with the emission of some energy.
Estimates based on the formulas of Ovichinnikov and
Schmid~\cite{ovi} lead to
 $\omega_{relax}\sim 100~kHZ$ which, as desired, would be much
slower
than the sweep time. This large difference between the tunneling
time and the relaxation time is due to  the  many
orders of magnitude suppression associated with the emission of
energy in the relaxation process, particularly for underdamped
SQUIDS.

 In conclusion, there are five important
 (inverse) timescales; $\omega_{0},~ \omega_{tunnel},~\epsilon,
~\omega_{sweep},~ D,$ and  $\omega_{relax}$. 
Naturally a wide range of behavior is available by changing the
parameters and some, like the sweep time or  the
temperature, can be adjusted online. 
  Probably the most important condition for 
 feasibility  is the smallness of 
$\omega_{relax}$: relaxation
processes should not be significant during the sweep and the
subsequent readout.

Concerning the readout itself,
 we have examined a scheme involving
  a switchable flux linkage to a DC SQUID,
 which would  be sufficiently fast for the above
estimates~\cite{dino}.
 Here one profits from the fact that observation
 of the system is only necessary after the procedure is completed.

\section{qbits and the quantum computer}  The two-state system
under discussion here suggests
itself as a physical embodiment of  the ``quantum computer''. 
  The ``qbit'' itself is naturally represented by
L and R playing the role of 0 and 1.  A  linear
combination of L and R may be created by adiabatically rotating
$\bf P$
from some starting position. 
 Adiabatic inversion is evidently an embodiment of NOT since it
 will turn one linear combination into another one with the weights
of L and R interchanged.
 As for  CNOT, the other basic operation,  a NOT is performed
 or not performed  on a
``target bit'' according to the state of a second, ``control
bit'', which itself does not change its state.
  One straightforward realization of this would be to
 perform the NOT operation as just described in the presence of an
additional linking flux
supplied by a  second  SQUID nearby.  The magnitude and direction
of this linking flux  would
be so arranged that the inversion of the first SQUID is or is not
successful depending on the state
of the second SQUID. This and many other interesting combinations
of junctions and flux
linkages may be contemplated and are under study~\cite{averin}.

SQUID systems like these would seem to be particularly well
suited  for the  embodiment of the quantum computer, where we
 wish to generate a series of unitary
transformations for  the various steps of computation.
 This may be done by creating 
a ``moving landscape" of potential maxima and minima, as in our
simplest
one-dimensional example of the adiabatic inversion. This imaginary
landscape 
  can be produced and manipulated by controlling various external
parameters
 (as with our sweeping flux), performing the various 
operations in one physical device. Naturally, practicality
will depend very much on the relation between the speed of these
operations and the decoherence/ relaxation times which we hope to
 determine by the present methods.

Although our interest here has been the SQUID, it will be evident
that the principle of determining decoherence times through the
inhibition of adiabatic inversion could be applied to many other
types of systems as well.

 Preliminary reports on  this work were presented by one of us
(P.S.)  at  ICSSUR VI (Napoli, Italy, 24-29 
May 1999), at NATO ASI  ( Ankara, 13-25 June 1999),
     and  at  MICRO AND NANO CRYOCENICS (Jyvaskyla, Finland, August
1-3 1999).


\end{document}